\documentstyle[multicol,prb,aps,epsf] {revtex}
\begin{document}
\bibliographystyle{prbsty}
\draft
\title{Localization length of a soliton from a non-magnetic impurity
       in a general double-spin-chain model}
\author{Tota Nakamura}
\address{Department of Applied Physics, Tohoku University,
         Sendai, Miyagi 980-8579, Japan }
\date{\today}
\maketitle
\begin {abstract}
A localization length of a free-spin soliton from a non-magnetic impurity
is deduced in a general double-spin-chain model 
($J_0$-$J_1$-$J_2$-$J_3$ model).
We have solved a variational problem 
which employs the nearest-neighbor singlet-dimer basis.
The wave function of a soliton is expressed by the Airy function, and
the localization length $(\xi)$ is found to obey a power law of 
the dimerization $(J_2-J_3)$ with an exponent $-1/3$;
$\xi\sim (J_2-J_3)^{-1/3}$.
This explains why NaV$_2$O$_5$ does not show the antiferromagnetic order,
while CuGeO$_3$ does by impurity doping.
When the gap exists by the bond-dimerization,
a soliton is localized and no order is expected.
Contrary, there is a possibility of the order
when the gap is mainly due to frustration.
\end  {abstract}
\begin{multicols}{2}
\narrowtext
A possibility of the superconductivity upon doping carriers to the
spin-ladder system has attracted much interest to the 
quasi-one dimensional quantum spin system these years.
\cite{dagotto9496}
Without the doping, the ground state of the spin-ladder
has strong singlet-dimer correlations along the rung pairs.
It leads to the finite energy gap and no long-range magnetic order.
Many intensive investigations were done in the course of doping 
impurities to the spin-gapped systems after this prediction.
As its byproduct, an unexpected anomaly was found.
That is, the long-range antiferromagnetic (AF) order appears
by doping non-magnetic impurities to a spin-gapped material.
\cite{azuma-htik97}
Copper sites of the base compound, SrCu$_2$O$_3$, is replaced with 
non-magnetic zinc atoms.
The impurity destroys the spin gap with only 1\% concentration, and
causes the AF order at low temperature.
This doping effect suddenly became a big topic both experimentally and
theoretically.
\cite{fujiwara-yfat98,fischer-llg98,miyazaki-touy97,laukamp-mgmd98,%
sorensen-aap98,augier-hdrs98}
The similar phenomena was also observed in inorganic spin-Peierls 
compound, CuGeO$_3$.\cite{hase-tsu93}
This compound can be explained theoretically by the one-dimensional
frustrated spin system with a ratio of the next-nearest-neighbor to
the nearest-neighbor interaction $\alpha\sim 0.35$.
\cite{fabricius-klbl97}
Thus, the origin of the gap is frustration.
On the other hand, another spin-Peierls compound NaV$_2$O$_5$ has not shown
the antiferromagnetic order by the Na depletion, 
while the energy gap disappear.
\cite{kuroe-sssiu98}
The origin of the gap of this system is the bond-dimerization.
\cite{isobe-u96}
We can attribute this difference to the origin of the gap.

Generally, one non-magnetic impurity atom destroys 
one singlet dimer pair, and makes one free spin-1/2.
We call it a soliton hereafter.
Solitons should interact with each other 
to be ordered antiferromagnetically.
Thus, a localization length of a soliton, $\xi$, 
should exceed a mean distance between impurities.
Up to now, an explicit expression for the wave function of a soliton 
or for its localization length has not given analytically.
Theoretical investigations have been done mainly by numerical methods.
\cite{miyazaki-touy97,laukamp-mgmd98,sorensen-aap98,augier-hdrs98}
In this paper, we deduce $\xi$ as a function of the strength of 
interaction bonds in a general double-spin-chain model by
using a variation which employs the nearest-neighbor singlet-dimer basis.
The result explains why a soliton localizes in the system whose
energy gap exists by the bond-dimerization.

We consider the following spin Hamiltonian as the base system under the open 
boundary conditions:
\begin{eqnarray}
 {\cal H} = \sum_{n=-N}^{N-1}
  ( &J_0& \mbox{\boldmath $\sigma$}_n \cdot\mbox{\boldmath $\sigma$}_{n+1}
  +  J_1  \mbox{\boldmath $\tau  $}_n \cdot\mbox{\boldmath $\tau  $}_{n+1}
  +  J_2  \mbox{\boldmath $\sigma$}_n \cdot\mbox{\boldmath $\tau  $}_{n}
\nonumber \\
  + &J_3& \mbox{\boldmath $\tau  $}_n \cdot\mbox{\boldmath $\sigma$}_{n+1}
  ) +J_2  \mbox{\boldmath $\sigma$}_N \cdot\mbox{\boldmath $\tau  $}_{N}
\label{eq:hamiltonian}
\end  {eqnarray}
\begin{figure}
    \epsfxsize = 8.0cm
\epsffile{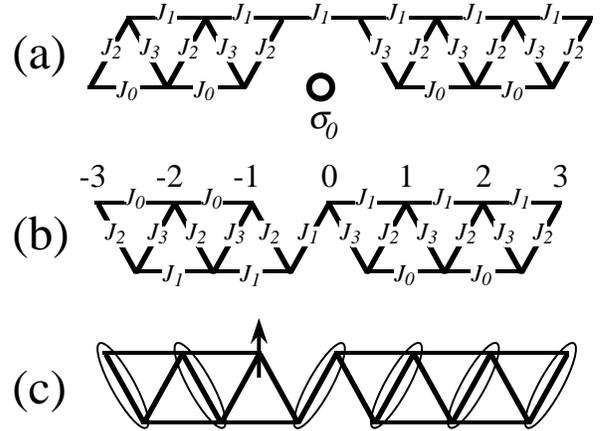}
 \caption {(a) Shape of the a general double spin-chain model with a
           single non-magnetic impurity at the origin,
            $\mbox{\boldmath $\sigma$}_0$.
           (b) An equivalent bond-depleted model 
               to the site-depleted model (a). 
               Soliton sites are indicated by numbers.
           (c) A schematic picture for the 
               variational basis $\psi_{i}$ for $i=-1$. An ellipse denotes a
               singlet-dimer state.
  \label{fig:lattice}
          }
\end  {figure}
Figure \ref{fig:lattice}(a) shows the shape of the lattice.
A non-magnetic impurity is located at the $\mbox{\boldmath $\sigma$}_0$-site.
This site-depleted lattice is equivalent to the bond-depleted lattice 
as shown in Fig. \ref{fig:lattice}(b), if we shift spins on the left 
side of the impurity by one lattice spacing to the right and exchange the
upper and the lower chain.
We make a variational analysis to this bond-depleted model.

The ground state of the base system has a strong
nearest-neighbor singlet-dimer correlation when it has a finite energy gap.
We consider this picture remains valid even after an impurity is doped, i.e.,
the ground state can be described in terms of one soliton and $2N$ 
nearest-neighbor singlet-dimer states.
Far-neighbor dimer states are neglected.
In this sense, the analysis is variational.
This soliton approach has already succeeded in explaining properties of the 
low-lying states in the fully-frustrated system.
\cite{tota-k96,tota-t97l,tota-t97}
The present approximation is good as long as the nearest-neighbor interactions,
$J_2$ and $J_3$ dominate the other bonds.

The variational basis is as depicted in Fig. \ref{fig:lattice} (c) and 
is written as follows.
\[
 \psi_i=[\sigma_{-N},\tau_{-N}]\cdots[\tau_{i-1},\sigma_{i}]
        \uparrow_i 
        [\tau_{i+1},\sigma_{i+1}]\cdots[\tau_{N},\sigma_{N}],
\]
where $i>0$, and
$[a,b]=(\uparrow_a\downarrow_b-\downarrow_a\uparrow_b)/\sqrt{2}$
denotes the singlet-dimer state. 
We convention that the spin-$a$ is always located in the upper chain,
and the spin-$b$ in the lower chain.
The basis $\psi_i$ for $i<0$ is defined in the same way.
A location of a soliton $\uparrow_i$ is restricted within the upper chain, 
thus there are $(2N+1)$ different states.
Note that
a state with a soliton in the lower chain can be expanded in a
linear combination of states with a soliton in the upper chain.
The present variational basis is not orthogonal to each other 
and satisfies the following relation:
\begin{eqnarray}
 \langle \psi_i  |\psi_j\rangle&=&\left ( \frac{1}{2}\right )^{|i-j|}.
                               \label{eq:overlap}
\end  {eqnarray}
The matrix elements of the Hamiltonian, 
$\langle \psi_i|{\cal H}|\psi_j\rangle$,
take different forms in regard to a sign of $i$ and $j$:
\begin{eqnarray}
 &&{ \langle \psi_i|{\cal H}|\psi_j\rangle}
     \Big/\left(-\frac{3}{4}\langle \psi_i  |\psi_j\rangle\right)\nonumber \\
  &=&-{\rm max}(i,j)(J_2-J_3) +|i-j|(J_2-J_1-J_0)\nonumber \\
    &&+2NJ_2 +J_0(1-\delta_{i,j}),\nonumber \\
    &&\mbox{~~~~~~~~ for ${\rm max}(i,j) \ge0$ and ${\rm min}(i,j) \ge 0$},\\
  &=& |i-j|(J_3-J_1-J_0)
      +2NJ_2 - J_3+2(J_1+J_0),\nonumber \\
    && \mbox{~~~~~~~~ for ${\rm max}(i,j) > 0$ and ${\rm min}(i,j) < 0$},\\
  &=& -{\rm min}(i,j)(J_3-J_1-J_0) 
      +2NJ_2+J_0+2J_1-J_3,\nonumber \\
    &&\mbox{~~~~~~~~ for ${\rm max}(i,j) = 0$ and ${\rm min}(i,j) < 0$},\\
  &=& {\rm max}(i,j)(J_2-J_3) +|i-j|(J_3-J_1-J_0)\nonumber \\
    &&+2NJ_2 +2J_1-J_3-J_1\delta_{i,j}, \nonumber \\
    &&\mbox{~~~~~~~~ for ${\rm max}(i,j) < 0$ and ${\rm min}(i,j) < 0$},
\end  {eqnarray}
where ${\rm max}(i,j)=i$ for $i>j$, ${\rm min}(i,j)=i$ for $i<j$, and
$\delta_{i,j}$ is the Kronecker delta.
The constant energy term is lower in the case of $i, j \ge 0$ than that 
in $i, j \le 0$, when $J_1 < J_3$.
This makes a soliton confined within the region of positive sites in the
low-energy limit.
Therefore, an impurity acts as a potential barrier to a soliton.

Our task is to find a function, $\Psi_{\rm var}=\sum_i C_i\psi_i$, that
minimizes the energy expectation,
\begin{equation}
 E_{\rm var}=\frac{\langle \Psi_{\rm var} |{\cal H}|
                 \Psi_{\rm var} \rangle}
        {\langle \Psi_{\rm var} |
                 \Psi_{\rm var} \rangle}
   = 
   \frac{\sum_{i,j}C_i C_j \langle \psi_i|{\cal H}|\psi_j\rangle}
        {\sum_{i,j}C_i C_j \langle \psi_i|         \psi_j\rangle}.
 \label {eq:vare}
\end  {equation}

If we diagonalize the denominator and rewrite the numerator with 
its eigenfunction,
this variational problem is transformed into a simple eigenvalue problem.
We first solve this transformed eigenvalue problem 
by the numerical diagonalization for a finite system with $N=200$.
After that, we give an
analytic solution in the continuous limit, $N\to\infty$,
when the wave number of the lowest energy state is zero.

\begin{figure}
    \epsfxsize = 8.0cm
\epsffile{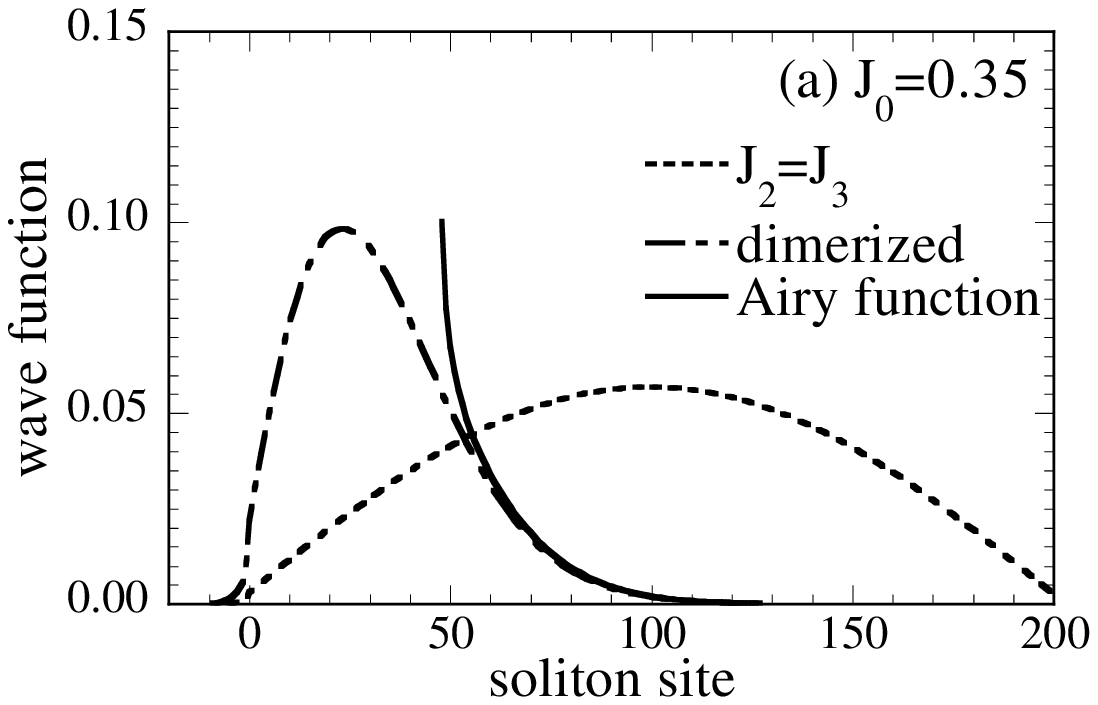}
    \epsfxsize = 8.0cm
\epsffile{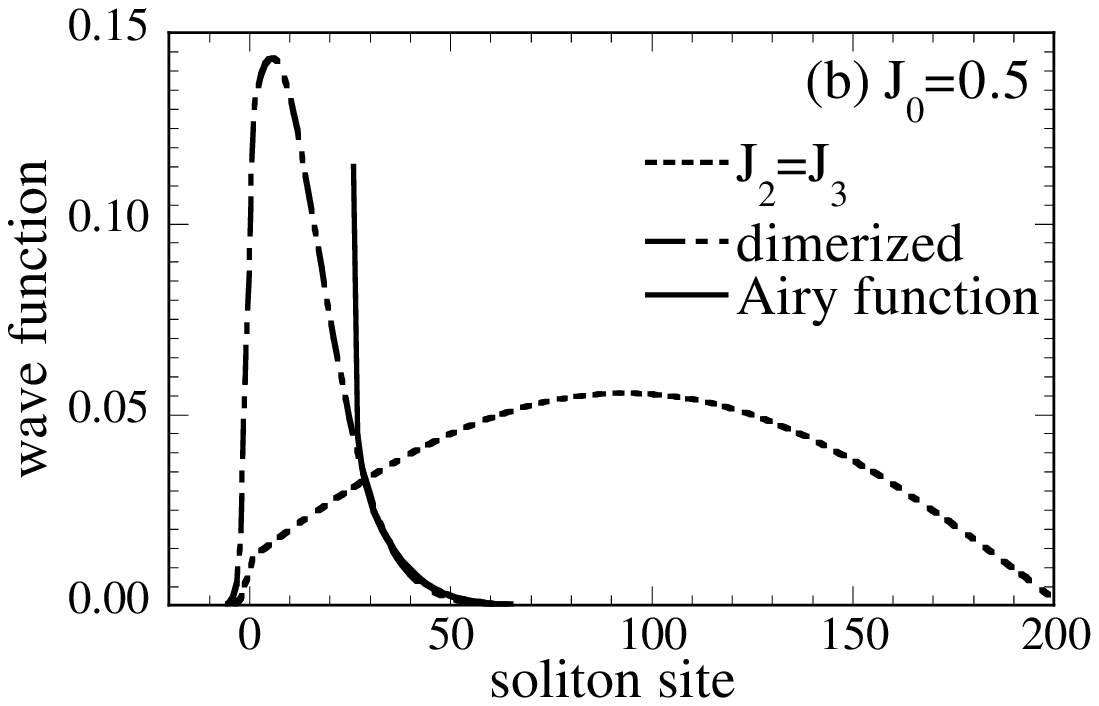}
    \epsfxsize = 8.0cm
\epsffile{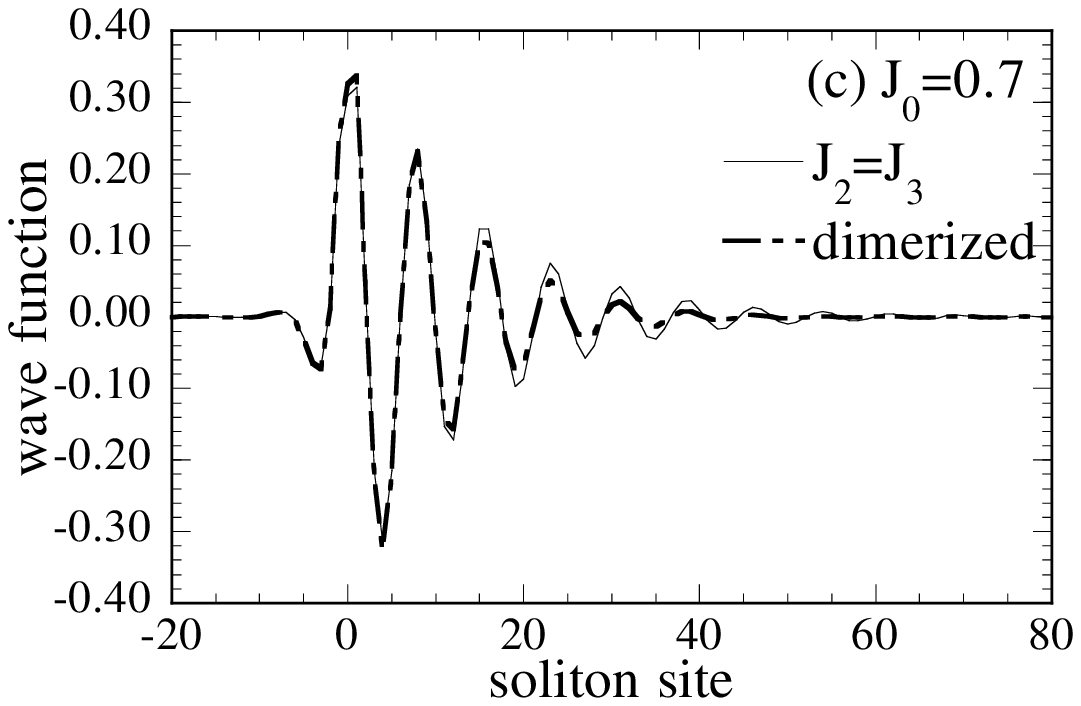}
 \caption {The ground-state wave function
           obtained by numerical diagonalization of the variational
           problem for (a):$J_0=0.35$, (b):$J_0=0.5$, and 
               (c):$J_0=0.7$. $J_3$ is set unity and $J_1=J_0$.
           The strength of the dimerization
           $J_2-J_3=2.441\times 10^{-4}$.
           Solid lines in (a) and (b) 
           are the asymptotic form of the Airy function.
  \label{fig:wave}
          }
\end  {figure}
Figure \ref{fig:wave}
shows behaviors of the ground-state wave function.
We set $J_3$ unity and vary $J_0=J_1$ as (a): $J_0=0.35$, which corresponds to
CuGeO$_3$; (b): $J_0=0.5$, the Majumdar-Ghosh model;
and (c): $J_0=0.7$, which is in the incommensurate phase.
Since our analysis is base on the nearest-neighbor singlet-dimer state, 
the results in the incommensurate phase is not beyond the speculation level.
For $J_0 \le 0.5$, the wave function of the non-dimerized case ($J_2=J_3$)
is the sine-function with a phase shift caused by the impurity.
The phase shift decreases with a decrease of $J_1$,
since the effective potential barrier at the impurity 
increases, and it becomes an open edge at $J_1=0$.
A soliton can be regarded as a free particle in a potential well
in this case.
When the dimerization is switched on, the wave function suddenly 
collapse to the impurity and exhibits an exponential decay.
The strength of the dimerization, $J_2-J_3$, is only $2.441\times 10^{-4}$
in this figure.
Therefore, a soliton becomes localized even with a sufficiently 
weak dimerization.
It can be noticed
that the wave function is more localized in the case $J_0=0.5$
as compared with that of $J_0=0.35$.
We will clarify this $J_0$-dependence of $\xi$ in the analytic solution.
In the case of $J_0 = 0.7$,
the wave function exhibits an oscillation with an incommensurate wave
number, $k_{\rm inc}$, and is already localized without the dimerization.
The dimerization effect only decreases a fraction of the localization length.
Since the origin of the localization in $J_0 \leq 0.5$ can be attributed
to the bond-dimerization alone,
the localization phenomenon in the incommensurate phase
is considered quite different from the former ones.
We leave to clarify this problem for the future investigation.

The variational problem can be solved analytically by the 
Fourier transformation, since the denominator is diagonalized in the
$N\to\infty$ limit.
Therefore, we rewrite the problem with a new basis,
$|\phi_k\rangle=\sum_n\exp[ikn]|\psi_n\rangle$.
The basis relation becomes
\begin{equation}
\langle\phi_k|\phi_l\rangle = \frac{3}{5-4\cos k}\delta_{k,l} + O(1/N).
\end  {equation}
The leading and the second terms of the diagonal part of the Hamiltonian, 
${\cal H}_{k,k}\equiv \langle\phi_k|{\cal H}|\phi_k\rangle$, are
\begin{eqnarray}
&&{\cal H}_{k,k}=\frac{9}{4} \cdot \frac{(J_2-J_3)}{5-4\cos k} \cdot
              \frac{N^2}{2N+1} 
+ \frac{3}{4}\left[\frac{3(J_2-J_0-2J_1)}{5-4\cos k} \right.\nonumber \\
&+& (J_0+J_1) 
 + \left. 8(J_2-J_1-J_0)\frac{4-5\cos k}{(5-4\cos k)^2}\right]
\frac{N}{2N+1},
\label{eq:kmatrix}
\end  {eqnarray}
where we have dropped the constant term, $2NJ_2$.
The leading term of the off-diagonal part, ${\cal H}_{k,l}$, is
\begin{eqnarray}
 &&-\frac{3}{4}\frac{(J_2-J_3)}{2N+1}\left[4N
    \frac{5\cos((k+l)N)-4\cos((k-l)N)}{(5-4\cos k)(5-4\cos l)}
    \right. \nonumber \\
 &+&\left. 
    \frac{3}{2}\left(\frac{1}{5-4\cos k}+\frac{1}{5-4\cos l}\right)
    \frac{1-\cos((k-l)N)}{1-\cos (k-l)}
    \right].
\end  {eqnarray}
The off-diagonal parts are always smaller than the diagonal parts by 
an order of $N$.

Before proceeding to the solution, we list several notices.
First, the wave number of the ground state must be zero, 
because we consider the continuous limit, $N\to\infty$ and $k\to 0$.
In the case of $J_0=J_1$, and $J_2=J_3$, the wave number that minimizes the
variational energy, $k_{\rm inc}$, is zero for $J_1/J_2 < 9/17$,
and otherwise it is given by
\begin{equation}
\cos k_{\rm inc}=\frac{1}{4}\left(5-\sqrt{9(2J_1-J2)/J_1}\right).
\end  {equation}
This incommensurate point was also obtained by the variational matrix-product 
{\it ansatz}.\cite{brehmer-kmn98}
Secondly, we find the diagonal part (\ref{eq:kmatrix}) diverge
for $J_2-J_3\neq 0$.
In order to avoid this divergence, we introduce a cutoff factor to the 
momentum as $k\to k+i\delta$, which has no effect to the physical results
by taking a limit $\delta\to 0$ after $N\to\infty$.

We rewrite the matrix elements and pick up only the leading term of 
the off-diagonal part and terms up to $k^2$ in the diagonal part.
Then, we have the continuous limit of the Hamiltonian $\tilde{\cal H}_{k,l}$:
\begin{eqnarray}
 \tilde{\cal H}_{k,l}
    &=&\left[{\rm const.}
      +\left( \frac{J_0+J_1}{4} +\frac{9}{2}(J_2-J_1-J_0) \right)k^2
                        \right]\delta_{k,l}\nonumber \\
     &-&\frac{3}{4}\frac{J_2-J_3}{2N+1}\left[
                   \frac{1}{(k-l-i\delta)^2}+\frac{1}{(k-l+i\delta)^2}\right],
\end  {eqnarray}
where the constant term is $(3J_3+6J_0+3J_1-8J_2)/8$.
Apart from this constant term, the Hamiltonian is equivalent to the 
following in the real space representation:
\begin{equation}
 {\cal H}_{\rm C} = -\frac{1}{2m}\frac{d^2}{dx^2} 
              + \frac{3}{4}(J_2-J_3)|x|\exp[-\delta |x|].
 \label{eq:continue}
\end  {equation}
Here, $x$ is the distance from the impurity, and $m$ is an effective mass of 
a soliton given by 
\begin{equation}
m^{-1}=(J_0+J_1)/2+9(J_2-J_1-J_0).
\label{eq:mass}
\end  {equation}
We take the limit $\delta\to 0$ at this stage.
The first term of the eq. (\ref{eq:continue})
is the kinetic energy of a soliton, and the second term
is the triangular potential which comes from the bond-dimerization.
By a scale transformation, $X=\theta x$, 
the eigenvalue equation ${\cal H}_c\Psi = E_{\rm C} \Psi$ becomes 
\begin{equation}
\left[-\frac{d^2}{dX^2}+(X-E')\right]\Psi = 0,
\end  {equation}
where $E'=E_{\rm C}\times 2m\theta^{-2}$ with $\theta^3=3m(J_2-J_3)/2$.
A solution of this equation
is known as the Airy function, $\Psi=Ai(X-E')$, with the first
eigenvalue $E'=2.338$.\cite{stern72} 
Then, the energy, $E_{\rm C}$, and the
localization length, $\xi$, of a soliton is estimated as
\begin{eqnarray}
E_{\rm C}=\frac{E'}{2m\theta^{-2}}&=&1.532\times m^{-1/3}(J_2-J_3)^{2/3},
\\
 \xi \sim (3+E')\times \theta^{-1}&=&4.663\times m^{-1/3}(J_2-J_3)^{-1/3}.
  \label{eq:xipower}
\end  {eqnarray}
The numeric factor, $(3+E')$, is determined since
the Airy function decays exponentially for large $X$ as 
\begin{equation}
A_i(X)\sim \frac{1}{2X^{1/4}}\exp [-2X^{3/2}/3],
\end  {equation}
and
$Ai(3)\sim 0.01$.
We plot this asymptotic form in Fig. \ref{fig:wave}(a) and (b)
and compare with the numerical solution.
The agreements are excellent,
even though we have neglected the impurity term of the off-diagonal
matrix elements in the analytic solution. 
As depicted in Fig. \ref{fig:lattice} (b), the $J_0$- and the $J_1$-bond
vanish and the $J_2$-bond changes to the $J_1$ at the impurity site.
This means the effective mass given by eq. (\ref{eq:mass})
is reduced to $m^{-1}=9J_1$ only at this site.
Therefore, the impurity attracts a soliton.
The analytic localization length, eq. (\ref{eq:xipower}),
should be modified as $\xi-a$ by a some constant term $a$ regarding
the phase shift.
We will determine $a$ by fitting $\xi-a$ to the numerical 
results of the variation in the following,
since the phase shift can not be obtained by the present analysis.

\begin{figure}
    \epsfxsize = 8.0cm
\epsffile{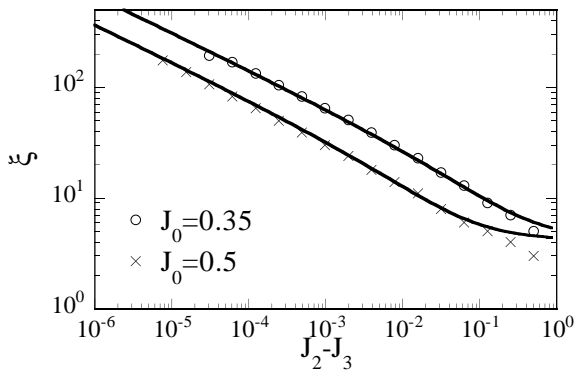}
 \caption {Localization length estimated from the wave function
           obtained by numerical diagonalization
           for $J_0=J_1=0.35$ ($\circ$) and $J_0=J_1=0.5$ ($\times$).
           $J_3$ is set unity. Lines are the analytic results of $\xi-a$
           with $a=5.5$.
  \label{fig:xi}
          }
\end  {figure}

Figure \ref{fig:xi} compares the analytic results of $\xi$ to 
those obtained by the numerical diagonalization.
The numerical data are estimated at the point where the absolute value of the 
normalized wave function $|\Psi_{\rm var}(\xi)|$ 
becomes smaller than $0.01$.
Analytic data depicted by lines are $\xi-a$ with $a=5.5$.
This small subtraction merely improves 
the fit in a small-$\xi$ region, $\xi\leq 20$.
It does not influence the asymptotic form of $\xi$
which is in a power law with an exponent $-1/3$.
We have checked this behavior remains in a wide region of
$0 < J_0/J_3 < 0.5$.
We can also predict the strength of the dimerization, $\delta=(J_2-J_3)/2$,
in CuGeO$_3$ by a plot of $J_0=0.35$ of this figure.
Since the AF order appears at the 2\% doping but 
disappears at the 1\%,\cite{hase-tsu93}
$\xi$ would be $12.5 < \xi < 25$, which roughly corresponds 
to $ 0.003 \leq \delta  \leq 0.025$.

In summary, 
we have given an explicit expression for the 
localization length of a soliton when a non-magnetic impurity is 
doped in a general double-spin-chain model.
The expression (\ref{eq:xipower}) classifies the behavior of a soliton 
by the origin of the energy gap.
When the gap originates in the bond-dimerization, a soliton is 
strongly localized near the impurity, which means there expects no
antiferromagnetic order.
On the other hand, a soliton can interact with another soliton 
when the gap is formed mainly by frustration, since $\xi$ diverges.
This result explains why NaV$_2$O$_5$ does not exhibit the 
AF order by the Na depletion, while CuGeO$_3$ does by the Zn doping.
The present analysis clarified that the motion of a soliton 
is well-described by a quantum particle in a potential well.
A soliton is confined to one side of the impurity by the potential
barrier proportional to $J_3-J_1$, and thus, it can
penetrate to the other side as $J_1$ approaches $J_3$.
It also feels a triangular potential
which is an outcome of the bond-dimerization.
We conclude that this triangular potential
is the cause of the localization of a soliton in the dimer phase.
Another type of the localization was also found 
in the incommensurate phase. 
It occurs irrespective of the bond-dimerization.
Derivation of the interaction between solitons and the localization 
problem in the incommensurate phase are left for the future investigations.

The author acknowledges thanks to Professor Hidetoshi Nishimori for the
diagonalization package Titpack Ver. 2.

   \end{multicols}
\end{document}